\documentclass[12pt]{article}
\usepackage{docmute}

\usepackage{epsf}
\usepackage{cite}
\usepackage{amsmath,amssymb}
\usepackage{pdfpages}
\usepackage{graphicx}
\usepackage{comment}
\usepackage{hyperref}
\usepackage{bm}
\usepackage{tikz}
\usetikzlibrary{positioning}
\usepackage{physics,cancel}

\definecolor{rossoferrari}{HTML}{D9073D}
\definecolor{mediumblue}{HTML}{0000CD}
\hypersetup{% hyperref option list
setpagesize=false,
bookmarksnumbered=true,%
bookmarksopen=true,%
colorlinks=true,%
linkcolor=rossoferrari,
urlcolor=mediumblue,
citecolor=mediumblue,
}

\newcommand{\mathhyphen}{\mathchar"2D}

\allowdisplaybreaks[1]

\newcommand{\lH}{\lambda_H}

\newcommand{\lL}{\lambda_{L}}

\newcommand{\lR}{\lambda_R}
\newcommand{\lLR}{\lambda_{LR}}

\newcommand{\lHL}{\lambda_{HL}}
\newcommand{\lHR}{\lambda_{HR}}

\newcommand{\sle}[1]{\tilde{\ell}_{#1}}
\newcommand{\smuon}[1]{\tilde{\mu}_{#1}}
\newcommand{\lam}[1]{\lambda_{#1}}

\newcommand{\Seff}{S_{\rm eff}}

\begin{document}

\renewcommand{\thefootnote}{\fnsymbol{footnote}}
\setcounter{footnote}{0}

\begin{titlepage}

\def\thefootnote{\fnsymbol{footnote}}

\begin{center}

\hfill KEK-TH-2404\\

\hfill March, 2022\\

\vskip .5in

{\Large \bf

  Upper bound on the smuon mass
  from vacuum stability
  in the light of muon $g-2$ anomaly

}

\vskip .5in

{\large
  So Chigusa$^{(a,b,c)}$, Takeo Moroi$^{(d)}$ and Yutaro Shoji$^{(e)}$
}

\vskip .5in

$^{(a)}$
{\em Berkeley Center for Theoretical Physics, Department of Physics,\\
University of California, Berkeley, CA 94720, USA}

\vskip 0.1in

$^{(b)}$
{\em Theoretical Physics Group, Lawrence Berkeley National Laboratory,\\
Berkeley, CA 94720, USA}

\vskip 0.1in

$^{(c)}$
{\em KEK Theory Center, IPNS, KEK, Tsukuba, Ibaraki 305-0801, Japan}

\vskip 0.1in

$^{(d)}$
{\em
Department of Physics, The University of Tokyo, Tokyo 113-0033, Japan
}

\vskip 0.1in

$^{(e)}$
{\em
Racah Institute of Physics, Hebrew University of Jerusalem, Jerusalem 91904, Israel
}

\end{center}
\vskip .5in

\begin{abstract}

  We derive an upper bound on the smuon mass assuming that the muon
  $g-2$ anomaly is explained by the supersymmetric (SUSY)
  contribution.  In the minimal SUSY standard model, the SUSY
  contribution to the muon $g-2$ is enhanced when the Higgsino mass
  parameter is large.  Then, the smuon-smuon-Higgs trilinear coupling
  is enhanced, which may destabilize the electroweak vacuum.  We
  calculate precisely the decay rate of the electroweak vacuum in such
  a case.  We include one-loop effects which are crucial to determine
  the overall normalization of the decay rate.  Requiring that the
  theoretical prediction of the muon anomalous magnetic moment is
  consistent with the observed value at the $1$ and $2\sigma$ levels
  (equal to the central value of the observed value), we found that
  the lightest smuon mass should be smaller than $1.38$ and $1.68\ {\rm
    TeV}$ ($1.20\ {\rm TeV}$) for $\tan\beta=10$ (with $\tan\beta$
  being the ratio of the vacuum expectation values of the two Higgs
  bosons), respectively, and the bound is insensitive to the value of
  $\tan\beta$.

\end{abstract}

\end{titlepage}

\renewcommand{\thepage}{\arabic{page}}
\setcounter{page}{1}
\renewcommand{\thefootnote}{\#\arabic{footnote}}
\setcounter{footnote}{0}
\renewcommand{\theequation}{\thesection.\arabic{equation}}

\section{Introduction}
\label{sec:intro}
\setcounter{equation}{0}

The muon $g-2$ measurements at the BNL and FermiLab
experiments had a great impact on the study of particle physics.  The value of the muon anomalous magnetic moment for these two combined is
\cite{Bennett:2002jb,Bennett:2004pv,Bennett:2006fi,Abi:2021gix}
\begin{align}
  a_\mu^{\rm (exp)} =
  (11\,659\,206.1 \pm 4.1 ) \times 10^{-10}.
  \label{amu(exp)}
\end{align}
On the contrary, the standard-model (SM) predicts
\cite{Aoyama:2020ynm}\footnote
{For more details about the estimation of the SM prediction, see
  Refs.\ \cite{Aoyama:2012wk, Aoyama:2019ryr, Czarnecki:2002nt,
    Gnendiger:2013pva, Davier:2017zfy, Keshavarzi:2018mgv,
    Colangelo:2018mtw, Hoferichter:2019mqg, Davier:2019can,
    Keshavarzi:2019abf, Kurz:2014wya, Melnikov:2003xd,
    Masjuan:2017tvw, Colangelo:2017fiz, Hoferichter:2018kwz,
    Gerardin:2019vio, Bijnens:2019ghy, Colangelo:2019uex,
    Blum:2019ugy, Colangelo:2014qya}.}
\begin{align}
  a_\mu^{\rm (SM)} =
  ( 11\,659\,181.0 \pm 4.3 ) \times 10^{-10}.
  \label{amu(SM)}
\end{align}
These values give
\begin{align}
  \Delta a_{\mu} \equiv a_\mu^{\rm (exp)} - a_\mu^{\rm (SM)}
  = ( 25.1 \pm 5.9) \times 10^{-10},
  \label{damu}
\end{align}
which shows $4.2\sigma$ discrepancy between the experimentally
measured value of $a_\mu$ and the SM prediction (the so-called muon $g-2$
anomaly). The discrepancy seems to strongly indicate the existence of a
physics beyond the SM (BSM), which can be the origin of the muon $g-2$
anomaly.

One of the attractive candidates of the BSM physics which can solve
the muon $g-2$ anomaly is the supersymmetry (SUSY).  In particular, in
the minimal SUSY SM (MSSM), the smuon-neutralino and
sneutrino-chargino diagrams may contribute significantly to the muon
anomalous magnetic moment
\cite{Lopez:1993vi,Chattopadhyay:1995ae,Moroi:1995yh}; the size of the
SUSY contribution can be as large as $\Delta a_{\mu}$ to solve the
muon $g-2$ anomaly (for the recent studies about the MSSM contribution
to the muon $g-2$, see, for example, \cite{Endo:2021zal,
  Chakraborti:2021dli, Han:2021ify, VanBeekveld:2021tgn,
  Ahmed:2021htr, Cox:2021nbo, Wang:2021bcx, Baum:2021qzx, Yin:2021mls,
  Iwamoto:2021aaf, Athron:2021iuf, Shafi:2021jcg, Aboubrahim:2021xfi,
  Chakraborti:2021bmv, Baer:2021aax, Aboubrahim:2021phn, Li:2021pnt,
  Jeong:2021qey, Ellis:2021zmg, Nakai:2021mha, Forster:2021vyz,
  Ellis:2021vpp, Chakraborti:2021mbr, Gomez:2022qrb,
  Chakraborti:2022vds, Agashe:2022uih}).  Because the superparticles
are in the loops, the SUSY contribution to the muon $g-2$ is
suppressed as the superparticles become heavy.  Thus, in order to
explain the muon $g-2$ anomaly, masses of (some of) the superparticles
are bounded from above.  A detailed understanding of the upper bound
is important in order to verify the SUSY interpretation of the muon
$g-2$ anomaly with ongoing and future collider experiments
\cite{Endo:2013lva, Endo:2013xka, Endo:2022qnm}.

The muon $g-2$ anomaly can be explained in various parameter regions
of the MSSM.  If the masses of all the superparticles are comparable,
the masses of superparticles are required to be of $O(100)\ {\rm
  GeV}$.  Then, the muon $g-2$ anomaly indicates that superparticles
(in particular, sleptons, charginos, and neutralinos) are important
targets of ongoing and future collider experiments.  The SUSY
contribution to the muon $g-2$ can be, however, sizable even if
superparticles are much heavier.  It happens when the Higgsino mass
parameter ({\it i.e.}, the so-called $\mu$ parameter) is significantly large
and the enlarged smuon-smuon-Higgs trilinear scalar coupling enhances the contribution to the muon $g-2$.
Such a trilinear coupling is, however, dangerous because it may make the
electroweak (EW) vacuum unstable \cite{Frere:1983ag, Gunion:1987qv,
  Casas:1995pd, Kusenko:1996jn}.

In this letter, we study the stability of the EW vacuum,
paying attention to the parameter region of the MSSM where the muon $g-2$ anomaly is solved (or alleviated) by the SUSY
contribution.  Requiring that the SUSY contribution to the muon anomalous
magnetic moment, denoted as $a_{\mu}^{\rm (SUSY)}$, be large enough
to solve the muon $g-2$ anomaly, the smuon masses are bounded from
above based on the observed longevity of the EW vacuum.
Refs.\ \cite{Endo:2013lva, Endo:2021zal} have studied the vacuum stability bound using a
tree-level analysis of the decay rate for the case where the SUSY
breaking mass parameters of the left- and right-handed sleptons are
degenerate.  The tree-level analysis, however, has several
inaccuracies.  In particular, it cannot fix the dimensionful prefactor
of the decay rate and receives the renormalization scale uncertainty.
These difficulties cannot be avoided without performing the
calculation at the one-loop level.  We study the stability of the
EW vacuum using the state-of-the-art method to calculate the
decay rate of the false vacuum \cite{Endo:2017gal, Endo:2017tsz,
  Chigusa:2020jbn}, with which a full one-loop calculation of the
decay rate is performed.  We also consider a wide range of the slepton
mass parameters.  Then, based on the accurate estimation of the decay
rate, we derive an upper bound on the lightest smuon mass to explain
the muon $g-2$ anomaly.

This letter is organized as follows.  In Section \ref{sec:mssm}, we
briefly overview the SUSY contribution to the muon anomalous magnetic
moment and discuss the importance of the stability of the EW
vacuum.  In Section \ref{sec:eft}, we introduce the effective field
theory (EFT) we use in our analysis.  In Section
\ref{sec:vacuumdecay}, we explain our procedure to calculate the decay
rate of the EW vacuum.  Our main results are given in Section
\ref{sec:results}.  Section \ref{sec:conclusions} is devoted for
conclusions and discussion.

\section{MSSM and muon $g-2$}
\label{sec:mssm}
\setcounter{equation}{0}

We first overview the model we consider, which is the low energy
effective theory obtained from the MSSM.  (For the review of the MSSM,
see, for example, Ref.\ \cite{Martin:1997ns}.)  We also explain why
the stability of the EW vacuum is important in the study of the SUSY
contribution to the muon $g-2$.

Since $a_{\mu}^{\rm (SUSY)}$ is enhanced in the
parameter region in which $\tan\beta$ is large \cite{Moroi:1995yh},
we concentrate on the large $\tan\beta$ case to obtain a conservative bound on the mass scale of the smuons.  Importantly, $\tan\beta$ cannot be arbitrarily
large if we require perturbativity.  In particular,
$\tan\beta$ is smaller than $\sim 50$ in the grand unified
theory (GUT), which is one of the strong motivations to consider the MSSM.
There, the coupling constants (in particular, the bottom Yukawa
coupling constant) should be perturbative up to the GUT scale.

\begin{figure}[t]
  \centering
  \includegraphics[width=0.65\linewidth]{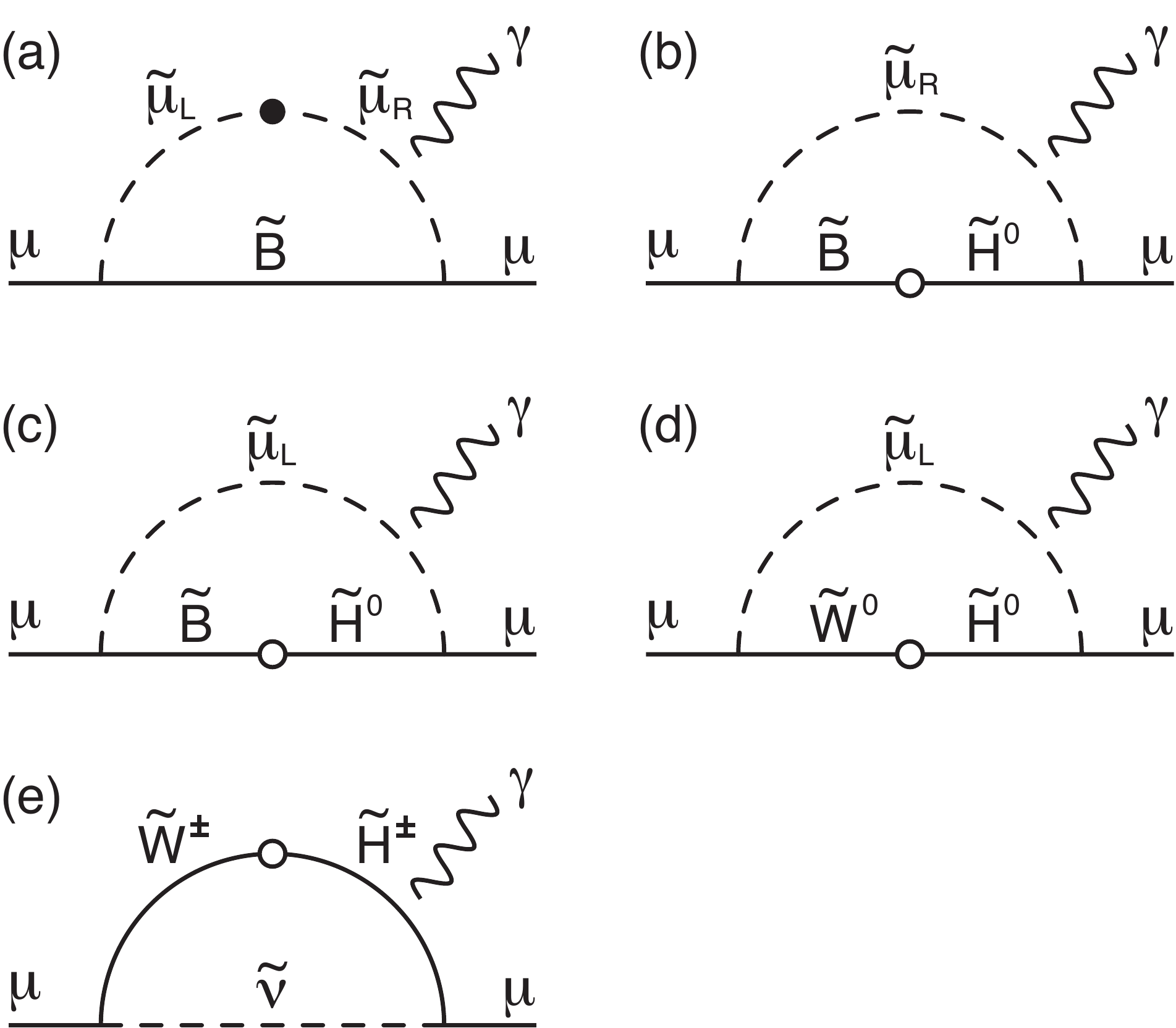}
  \caption{One-loop Feynman diagrams, which are enhanced in large
    $\tan\beta$ limit, giving rise to the SUSY contribution to the
    muon $g-2$.  Here, the mass insertion approximation is adopted.
    The black and white blobs are two-point interactions induced by
    the VEVs of Higgs bosons.}
  \label{fig:feyndiags}
\end{figure}

In order to study the behavior of $a_{\mu}^{\rm (SUSY)}$ in the large
$\tan\beta$ case, it is instructive to use the so-called mass
insertion approximation in which $a_{\mu}^{\rm (SUSY)}$ is estimated
in the gauge-eigenstate basis and the interactions
proportional to the Higgs vacuum expectation values (VEVs) are treated as
perturbations.  (In our following numerical calculation, $a_{\mu}^{\rm
  (SUSY)}$ is estimated more precisely by using the basis in which the
sleptons, charginos, and neutralinos are in the mass eigenstates, as
we will explain.)  In Fig.\ \ref{fig:feyndiags}, we show one-loop
diagrams which may dominate the SUSY contribution to the muon $g-2$ in
the large $\tan\beta$ limit.  Because the superparticles are in the
loop, $a_{\mu}^{\rm (SUSY)}$ is suppressed as the superparticles
become heavier.  For the case where the masses of all the
superparticles are comparable, for example, the SUSY contribution to
the muon anomalous magnetic moment is approximately given by
$|a_\mu^{\rm (SUSY)}|\simeq \frac{5g_2^2}{192\pi^2}
\frac{m_\mu^2}{m_{\rm SUSY}^2}\tan\beta$, where $g_2$ is the gauge
coupling constant of $SU(2)_L$, $m_\mu$ is the muon mass, and $m_{\rm
  SUSY}^2$ is the mass scale of superparticles.  (Here, the
contributions of the diagrams that contain the Bino are neglected
because they are subdominant.)  Taking $\tan\beta\sim 50$, which is
the approximate maximum possible value of $\tan\beta$ for the
perturbativity up to the GUT scale, the superparticles should be
lighter than $\sim 700\ {\rm GeV}$ in order to make the total muon
anomalous magnetic moment consistent with the observed value at the
$2\sigma$ level.

Such an upper bound is significantly altered by the Bino-smuon diagram
(Fig.\ \ref{fig:feyndiags}~(a)).  The other diagrams ({\it i.e.},
Fig.\ \ref{fig:feyndiags}~(b) $-$ (e)) have slepton, gaugino, and
Higgsino propagators in the loop, and hence their contributions are
suppressed when any of these particles is heavy.  On the contrary, the
Bino-smuon diagram has only the smuon and Bino propagators in the
loop, and its contribution is approximately proportional to the
Higgsino mass parameter $\mu$.  Thus, with a very large $\mu$ parameter,
the contribution of the Bino-smuon diagram can be large enough to cure
the muon $g-2$ anomaly even if the smuon and/or Bino are much
heavier than the upper bound estimated above.

In the following, we study the upper bound on the masses of
superparticles in the light of the muon $g-2$ anomaly, paying
particular attention to the contribution of the Bino-smuon diagram.
In the parameter region where the Bino-smuon diagram
has a dominant contribution, a large $\mu$ parameter enhances the
smuon-smuon-Higgs trilinear coupling.  Such a large trilinear scalar
coupling is dangerous because it may destabilize the EW
vacuum.  Consequently, the lifetime of the EW vacuum may
become shorter than the present cosmic age
\cite{ParticleDataGroup:2020ssz}:
\begin{align}
  t_{\rm now} \simeq 13.8\ {\rm Gyr}.
  \label{t_now}
\end{align}
The parameter region predicting the too short lifetime of the EW
vacuum is excluded.

The purpose of this letter is to derive an upper bound on the smuon mass under
the requirement that the muon $g-2$ anomaly be solved (or relaxed) by
the SUSY contribution.  We are interested in the case where the $\mu$
parameter is large so that the Bino-smuon diagram dominates
$a_\mu^{\rm (SUSY)}$; hereafter, we consider the case where $\mu$ is
much larger than the Bino and smuon masses.  A large value of $\mu$
implies heavy Higgsinos.  In addition, the stops are expected to be relatively heavy to push up the lightest Higgs mass to the observed value,
{\it i.e.}, about $125\ {\rm GeV}$, through the radiative correction
\cite{Okada:1990vk, Okada:1990gg, Ellis:1990nz, Haber:1990aw}.  On the
contrary, in order to enhance $a_\mu^{\rm (SUSY)}$, slepton and Bino
masses should be close to the EW scale.  Based on these
considerations, in this letter, we consider the case where the Bino
$\tilde{B}$ and smuons are relatively light among the MSSM particles.
These particles are assumed to have EW-scale masses comparable to the
top pole mass $M_t$.  The other MSSM constituents are assumed to have
heavier masses, which are characterized by a single scale $M_S$.  (For
simplicity, masses of gauginos other than $\tilde{B}$ are assumed to
be of $O(M_S)$.)  We assume that there exists a significant hierarchy
between $M_t$ and $M_S$ and thus the superparticles with the masses of
$\sim M_S$ do not affect the physics of our interest.  A comment on
the case where some of the other SUSY particles are as light as the smuons
will be given at the end of this letter.

A large value of $\mu$ suggests relatively large values of the soft
SUSY breaking Higgs mass parameters for a viable EW symmetry
breaking; the light Higgs mass ({\it i.e.}, the mass of the SM-like Higgs
boson) is realized by the cancellation between the contributions of
the $\mu$ and soft SUSY breaking parameters.  The heavier Higgs doublet is
expected to have masses of $O(M_S)$ which is comparable to $\mu$.  In
such a case, the SM-like Higgs doublet, denoted as $H$, and the heavier doublet,
$H'$, are given by linear combinations of the up- and down-type
Higgs bosons, denoted as $H_u$ and $H_d$, respectively, as
\begin{align}
  \left( \begin{array}{c} H \\ H' \end{array} \right)
  =
  \left( \begin{array}{cc}
    \cos\beta & \sin\beta
    \\
    -\sin\beta & \cos\beta
  \end{array} \right)
  \left( \begin{array}{c} H_d\\ H_u \end{array} \right),
\end{align}
where $\tan\beta$ is the ratio of the vacuum expectation values (VEVs)
of up- and down-type Higgs bosons.

In the case of our interest, the mass spectrum around the EW scale
includes the second-generation sleptons and Bino $\tilde{B}$, as well
as the SM particles.  Hereafter, the sleptons in the second generation
in the gauge eigenstate are denoted as $\sle{L}$ and
$\smuon{R}$; $\sle{L}$ is an $SU(2)_L$ doublet with
hypercharge $\frac{1}{2}$, which is decomposed as
\begin{align}
  \sle{L} =
  \begin{pmatrix}
    \tilde{\nu}_{L} \\ \smuon{L}
  \end{pmatrix},
\end{align}
while $\smuon{R}$ is an $SU(2)_L$ singlet with
hypercharge $-1$.

\section{Effective field theory analysis}
\label{sec:eft}
\setcounter{equation}{0}

We are interested in the case where there exists a hierarchy in the
mass spectrum of the MSSM particles.  To deal with the hierarchy, we
resort to the EFT approach and solve the renormalization group (RG)
equations with proper boundary conditions to evaluate the EFT coupling
constants.  Hereafter, we assume that the effects of possible
CP-violating phases are negligible.

We adopt $M_t$ and $M_S$ as matching scales.  For the renormalization
scale $Q < M_t$ (with $Q$ being the renormalization scale), we
consider the QCD+QED that contains the SM gauge couplings and fermion
masses as parameters.  For $M_t < Q < M_S$, we consider an EFT with
Bino and smuons as described below.  At $Q=M_S$, the EFT is matched to
the full MSSM, which imposes relations among EFT couplings.  We choose
$M_S$ to be close to the Higgsino mass.

The Lagrangian of the EFT, which is relevant for the calculation of
the decay rate of the EW vacuum and the muon $g-2$, is given by
\begin{align}
  \mathcal{L} = \mathcal{L}_{\mathrm{SM}}
  + \Delta \mathcal{L}_{\mathrm{kin}}
  + \Delta \mathcal{L}_{\mathrm{mass}}
  + \Delta \mathcal{L}_{\mathrm{Yukawa}} - V,
\end{align}
where $\mathcal{L}_{\mathrm{SM}}$ is the SM Lagrangian without the
Higgs potential, and the additional kinetic terms, mass terms, and
Yukawa couplings are described by
\begin{align}
  \Delta \mathcal{L}_{\mathrm{kin}} =& \,
  | D_\mu \sle{L} |^2 + | D_\mu \smuon{R}|^2
  - i \tilde{B} \sigma^\mu \partial_\mu \tilde{B}^\dagger, \\
  \Delta \mathcal{L}_{\mathrm{mass}} =& \,
  - \frac{1}{2} M_1 \tilde{B}\tilde{B} + \mathrm{h.c.}, \\
  \Delta \mathcal{L}_{\mathrm{Yukawa}} =& \,
  Y_{L} \sle{L}^\dagger \ell_{L} \tilde{B}
  + Y_{R} \smuon{R}^\dagger \mu_R \tilde{B}^\dagger
  + \mathrm{h.c.},
\end{align}
where $\ell_L$ and $\mu_R$ are the second generation left-handed lepton doublet and right-handed lepton, respectively.
We use the two-component Weyl notation for fermions.
The scalar potential $V$ is given by
\begin{align}
  V = &\,  V_2 + V_3 + V_4,
  \label{Vtot}
\end{align}
with
\begin{align}
  V_2 = &\,  m_H^2 |H|^2
  + m_{L}^2\, | \sle{L} |^2
  + m_{R}^2\, | \smuon{R} |^2,
  \label{eq:V2} \\
  V_3 = &\,  - T H^\dagger \sle{L} \smuon{R}^\dagger + \text{h.c.},\\
  V_4 = &\,  \lambda_H |H|^4
  + \lam{HL} |H|^2 | \sle{L} |^2
  + \lam{HR} |H|^2 | \smuon{R} |^2
  + \kappa
  ( H^\dagger \sle{L} ) ( \sle{L}^\dagger H )
  \nonumber \\ &\,
  + \lam{L} | \sle{L} |^4
  + \lam{R} | \smuon{R} |^4
  + \lam{LR} | \sle{L}| ^2 | \smuon{R} |^2,
  \label{eq:V4}
\end{align}
where $T$ is the trilinear scalar coupling constant.

Next, we describe the matching conditions of coupling constants at the
threshold scales.  All the SM parameters including the Higgs quartic
coupling ${\lambda}_H^{\rm (SM)}$ and the mass squared parameter
$m_H^{2{\rm (SM)}}$ are determined at $Q=M_t$.  Importantly, the top
Yukawa coupling, the gauge couplings, the Higgs quartic coupling, and
the Higgs mass parameter are subject to the possibly large weak-scale
threshold corrections.  We use the results of \cite{Buttazzo:2013uya}
to fix these parameters with using physical parameters
$\alpha_3(M_Z)=0.1179$, $M_t=172.76\,\mathrm{GeV}$,
$M_W=80.379,\mathrm{GeV}$, and $M_h=125.25\,\mathrm{GeV}$
\cite{ParticleDataGroup:2020ssz}.  As for the light fermion couplings,
we calculate the running of their masses with the one-loop QED and
three-loop QCD beta functions \cite{Gorishnii:1990zu, Tarasov:1980au,
  Gorishnii:1983zi} to determine the corresponding Yukawa couplings at
$Q=M_t$.

For other parameters, we mostly adopt the tree-level matching between
the SM and the EFT at $Q=M_t$, but take into account some of the
one-loop corrections which can be sizable.  The corrections to the
Higgs quartic coupling and the mass term are given by
\begin{align}
  \lH &= \lambda_H^{\rm (SM)} + \Delta \lH, \label{eq:dellH} \\
  m_H^2 &= m_H^{2{\rm (SM)}} + \Delta m_H^2, \label{eq:delmHSq}
\end{align}
with
\begin{align}
  (16\pi)^2 \Delta \lH =& \left(
  \lam{HL}^2 + \lam{HL} \kappa + \frac{1}{2} \kappa^2
  \right) B_0(m_{L}^2, m_{L}^2)
  + \frac{1}{2} \lam{HR}^2 B_0(m_{R}^2, m_{R}^2) \notag \\
  &+ (\lam{HL}+\kappa) T^2 C_0 (m_{L}^2, m_{L}^2, m_{R}^2)
  + \lam{HR} T^2 C_0 (m_{R}^2, m_{R}^2, m_{L}^2) \notag \\
  &+ \frac{1}{2} T^4 D_0(m_{L}^2, m_{R}^2, m_{L}^2, m_{R}^2),\\
  (16\pi)^2 \Delta m_H^2 =&
  \left( 2\lam{HL} + \kappa \right) A_0(m_{L}^2)
  + \lam{HR} A_0(m_{R}^2)
  + T^2 B_0 (m_{L}^2, m_{R}^2),
\end{align}
where $A_0$, $B_0$, $C_0$, and $D_0$ are the Passarino-Veltman one-,
two-, three-, and four-point functions without momentum inflow,
respectively \cite{Passarino:1978jh}.  In determining the muon Yukawa
coupling in the EFT, we also take account of the one-loop correction
\cite{Marchetti:2008hw, Girrbach:2009uy} because it may significantly
affect the vacuum decay rate and $a_{\mu}^{\rm (SUSY)}$.  The
correction $\Delta y_\mu$ is given by
\begin{align}
  (16\pi)^2 \Delta y_{\mu} = Y_{L} Y_{R} T M_1
  J(M_1^2, m_{R}^2, m_{L}^2),
  \label{dymu}
\end{align}
with
\begin{align}
  J(a,b,c) \equiv -\frac
  {ab\ln(a/b) + bc\ln(b/c) + ca\ln(c/a)}
  {(a-b)(b-c)(c-a)}. \label{eq:I}
\end{align}
The muon Yukawa coupling constant in the EFT, $y_\mu$, and that in the
SM, $y_\mu^{\rm (SM)}$, are related as $y_\mu=y_\mu^{\rm (SM)}+\Delta
y_{\mu}$.  In the present case, the sign of $\Delta y_{\mu}$ is
correlated with that of $a_\mu^{\rm (SUSY)}$ and is negative.  These
corrections can be sizable due to the hierarchy of the scales
$M_1,\,m_L,\,m_R \ll M_S$.  Concerning the trilinear coupling $T$, the
value is determined so that an input value of $a_\mu^{\rm (SUSY)}$ is
realized.

In the MSSM, the scalar potential is completely determined only by the parameters of the superpotential, {\it i.e.}, the Yukawa and gauge couplings.
Accordingly, we impose the matching conditions on the EFT couplings at the matching scale $Q=M_S$.
At the tree level, these conditions are given by
\begin{align}
  Y_L = \frac{1}{\sqrt{2}} g_Y = \sqrt{\frac{3}{10}} g_1, ~~~
  Y_R = -\sqrt{2} g_Y = -\sqrt{\frac{6}{5}} g_1,
\end{align}
where $g_Y$ and $g_1$ are the $U(1)_Y$ gauge coupling constant and its
$SU(5)$-normalized value, respectively, and
\begin{align}
  \lR &= \frac{3}{10} g_1^2,\\
  \lL &= \frac{1}{8} g_2^2 + \frac{3}{40} g_1^2,\\
  \lLR &= \frac{y_\mu^2}{\cos^2 \beta} - \frac{3}{10} g_1^2, \label{eq:lLR} \\
  \lHR &= y_\mu^2 - \frac{3}{10} g_1^2 \cos 2\beta, \label{eq:lHRm} \\
  \lHL &= \left( \frac{1}{4} g_2^2 + \frac{3}{20} g_1^2 \right) \cos 2\beta,\\
  \kappa &= y_\mu^2 - \frac{1}{2} g_2^2 \cos 2\beta. \label{eq:lkappam}
\end{align}
The $T$-parameter is related to the MSSM parameters as
\begin{align}
  T &= y_\mu \mu \tan\beta + A_\mu \cos\beta,
  \label{T-param}
\end{align}
with $A_\mu$ being the soft SUSY breaking trilinear scalar coupling of
smuon.  For simplicity, we assume that the SUSY breaking contribution
to the scalar trilinear coupling, $T$, is negligible.  We expect that
this assumption is valid when $\mu\tan\beta$ is much larger than the
typical smuon masses, which is the case in our following discussion.
Notice that, as we see below, the SUSY contribution to the muon $g-2$
and the decay rate of the EW vacuum are both dependent on the MSSM
parameters through the $T$-parameter.  Thus, the upper bound on the
smuon mass, which will be derived in the following Sections, will be
almost unchanged even if the effect of $A_\mu$ on $T$ is sizable.
Notice that we use Eq.\ \eqref{T-param} only to evaluate $\mu$.

Although we determine $\lH$ at $Q=M_t$, there is also the SUSY
relation between $\lH$ and other couplings.  Considering only the stop
contribution to the threshold correction at $Q=M_S$, we
obtain the one-loop matching condition \cite{Bagnaschi:2014rsa},
\begin{align}
  \lambda_H =
  \left( \frac{1}{8} g_2^2 + \frac{3}{40} g_1^2 \right) \cos^2 2\beta
  + \delta \lambda_H,
  \label{eq:lambdaH_matching}
\end{align}
with
\begin{align}
  (16\pi^2) \delta \lambda_H \simeq &\, \frac{3}{2} y_t^2 \left[ y_t^2 + \left( \frac{1}{2} g_2^2 - \frac{1}{10} g_1^2 \right) \cos2 \beta \right]
  \ln \frac{m_{Q3}^2}{Q^2} \notag \\ &\,
+ \frac{3}{2} y_t^2 \left( y_t^2 + \frac{2}{5} g_1^2 \cos2 \beta \right)
  \ln \frac{m_{U3}^2}{Q^2} \notag \\ &\,
+ \frac{\cos^2 2\beta}{200} \left[ (25g_2^4 + g_1^4) \ln \frac{m_{Q3}^2}{Q^2}
  + 8 g_1^4 \ln \frac{m_{U3}^2}{Q^2}
  + 2 g_1^4 \ln \frac{m_{D3}^2}{Q^2} \right],
\end{align}
where $y_t$ is the top-quark Yukawa coupling constant while $m_{Q3}$,
$m_{U3}$, and $m_{D3}$ are the mass parameters of the third generation
left-handed squark, right-handed up-type squark, and right-handed
down-type squark, respectively.  For simplicity, we neglect the
threshold correction to the parameters of the superpotential, {\it i.e.},
the top Yukawa coupling and the gauge couplings, and use their values
in the EFT at $Q=M_S$ to evaluate the size of $\delta \lambda_H$.
Once the value of $\lH$ at the matching scale $M_S$ is obtained, we
can solve \eqref{eq:lambdaH_matching} against the stop mass
$m_{\tilde{t}}$ assuming the universality $m_{\tilde{t}} \equiv m_{Q3}
= m_{U3} = m_{D3}$.
Requiring the observed Higgs mass to be realized, we have checked that
difference between $|\mu|$ and $m_{\tilde{t}}$ is within one or two
orders of magnitude in the region with small enough decay rate of the
EW vacuum.\footnote
{In some case, $m_{\tilde{t}}$ becomes one or two orders of
  magnitude smaller than $|\mu|$ and it may induce a color breaking
  minimum where stops acquire VEVs.  We do not discuss the instability
  due to such a color breaking minimum because it depends on various
  fields and parameters that are not included in our EFT.}

For $M_t < Q < M_S$, we solve the RG equations of the EFT.
We use the two-loop RG equations \cite{Luo:2002ey} augmented by some important three-loop contributions calculated in \cite{Buttazzo:2013uya} for the SM-like couplings.
On the other hand, Bino and smuon contributions to the beta functions of the SM-like couplings and the beta functions of the couplings specific to the EFT are calculated at the one-loop level.
Since all the SM parameters are fixed at $Q=M_t$ and below, while the other couplings are determined at $Q=M_S$, we iteratively solve the RG evolution in $M_t < Q < M_S$ to obtain consistent solutions.

Next, we explain how we calculate the SUSY contribution to the muon
$g-2$.  Because we are interested in the case where the masses of Bino
and smuons are much lighter than Higgsino (and other superparticles),
the EFT parameters introduced above are used.

The mass matrix of the smuons is given by
\begin{align}
  {\bf M}^2_{\tilde{\mu}} =
  \left( \begin{array}{cc}
    m_L^2 + (\lambda_{HL}+\kappa) v^2 & - T v
    \\
    - T v & m_R^2 + \lambda_{HR} v^2
  \end{array} \right),
\end{align}
where $v\simeq 174\ {\rm GeV}$ is the vacuum expectation value of the
SM-like Higgs.  The mass matrix can be diagonalized by a
$2\times 2$ unitary matrix $U$ as
\begin{align}
  \mbox{diag} (m^2_{\tilde{\mu}_1}, m^2_{\tilde{\mu}_2}) =
  U^\dagger {\bf M}^2_{\tilde{\mu}} U,
\end{align}
and the gauge eigenstates are related to the mass eigenstates, denoted
as $\tilde{\mu}_A$ ($A=1$, $2$), as
\begin{align}
  \left( \begin{array}{c}
    \tilde{\mu}_L \\ \tilde{\mu}_R
  \end{array} \right) =
  U \left( \begin{array}{c}
    \tilde{\mu}_1 \\
    \tilde{\mu}_2
  \end{array} \right) \equiv
  \left( \begin{array}{cc}
    U_{L,1} & U_{L,2}
    \\
    U_{R,1} & U_{R,2}
  \end{array} \right)
  \left( \begin{array}{c}
    \tilde{\mu}_1 \\ \tilde{\mu}_2
  \end{array} \right).
\end{align}

At the one-loop level, the Bino-smuon loop contributions to the muon
anomalous magnetic moment is given by \cite{Moroi:1995yh}
\begin{align}
  a_\mu^{({\rm SUSY},\, 1\mathhyphen{\rm loop})} =
  \frac{m_\mu^2}{16\pi^2}
  \sum_{A=1}^2
  \frac{1}{m_{\tilde{\mu}_A}^2}
  \left[
    - \frac{1}{12} \mathcal{A}_A f_1 (x_A)
    - \frac{1}{3} \mathcal{B}_A f_2 (x_A)
    \right],
\end{align}
where $x_A\equiv M_1^2/m_{\tilde{\mu}_A}^2$,
\begin{align}
  \mathcal{A}_A \equiv Y_L^2 U_{L,A}^2 + Y_R^2 U_{R,A}^2,~~~
  \mathcal{B}_A \equiv \frac{M_1 Y_L Y_R U_{L,A} U_{R,A}}{m_\mu},
\end{align}
and the loop functions are given by
\begin{align}
  f_1 (x) \equiv &\,
  \frac{2}{(1-x)^4} (1 - 6x + 3x^2 + 2x^3 - 6x^2 \ln x),
  \\
  f_2 (x) \equiv &\, \frac{3}{(1-x)^3} (1 - x^2 + 2 x \ln x).
\end{align}

In the MSSM, some of the two-loop contributions to the muon anomalous
magnetic moment may become sizable.  One important contribution is
the non-holomorphic correction to the muon Yukawa coupling
constant \cite{Marchetti:2008hw, Girrbach:2009uy}.  In the limit of
large $\tan\beta$ (or, large $T$), such an effect can be
significant.  In the present setup, such a non-holomorphic correction
to the muon Yukawa coupling constant is taken into account when the
EFT parameters (in particular, $y_\mu$) are matched to the MSSM
parameters at the SUSY scale.  Another is the photonic two-loop
correction \cite{Degrassi:1998es, vonWeitershausen:2010zr}.  Such a
contribution includes large QED logarithms and can affect the SUSY
contribution to the muon $g-2$ by $\sim 10\ \%$ or more.  The full
photonic two-loop correction relevant for our analysis is given by
\cite{vonWeitershausen:2010zr}
\begin{align}
  a_\mu^{({\rm SUSY,\, photonic})} = & \,
  \frac{m_\mu^2}{16\pi^2} \frac{\alpha}{4\pi}
  \sum_{A=1}^2 \frac{1}{m_{\tilde{\mu}_A}^2}
  \Bigg[
    16
    \left\{
    - \frac{1}{12} \mathcal{A}_A f_1 (x_A)
    - \frac{1}{3} \mathcal{B}_A f_2 (x_A)
    \right\} \ln \frac{m_\mu}{m_{\tilde{\mu}_A}}
    \nonumber \\ & \,
    - \left\{
    - \frac{35}{75} \mathcal{A}_A f_3 (x_A)
    - \frac{16}{9} \mathcal{B}_A f_4 (x_A)
    \right\}
    + \frac{1}{4} \mathcal{A}_A f_1 (x_A)
    \ln \frac{m_{\tilde{\mu}_A}^2}{Q_{\rm DREG}^2}
    \Bigg],
\end{align}
where $\alpha$ is the fine structure constant, $Q_{\rm DREG}$ is the
dimensional-regularization scale, and
\begin{align}
  f_3 (x) \equiv &\, \frac{4}{105(1-x)^4}
  [
    (1-x) (-97x^2 -529x +2)
    + 6 x^2 (13x + 81) \ln x
    \nonumber \\ &\,
    +108x (7x + 4) \mbox{Li}_2 (1-x)
  ],
  \\
  f_4 (x) \equiv &\, \frac{-9}{4(1-x)^3}
  [
    (x+3) (x \ln x +x -1)
    + (6x+2) \mbox{Li}_2 (1-x)
  ].
\end{align}

In our analysis, the SUSY contribution to the muon anomalous magnetic
moment is evaluated as
\begin{align}
  a_\mu^{\rm (SUSY)} =
  a_\mu^{({\rm SUSY},\, 1\mathhyphen{\rm loop})} +
  a_\mu^{({\rm SUSY,\, photonic})},
\end{align}
using the EFT parameters evaluated at the renormalization scale $Q=M_t$.
We note that the above prescription gives a good estimation of the
SUSY contribution to the muon anomalous magnetic moment in the
parameter region we consider in the following discussion.  In
particular, for the case of our interest, the effect of the
Bino-Higgsino-smuon diagrams ({\it i.e.}, Fig.\ \ref{fig:feyndiags} (b) and
(c)) is estimated to be $O(0.1)\ \%$ or smaller relative to
$a_\mu^{\rm (SUSY)}$ given above.  The Wino-Higgsino-slepton diagrams
({\it i.e.}, Fig.\ \ref{fig:feyndiags} (d) and (e)) become irrelevant in the
decoupling limit of the Winos.

\section{Decay rate of electroweak vacuum}
\label{sec:vacuumdecay}
\setcounter{equation}{0}

Using the method proposed by Callan and Coleman \cite{Coleman:1977py,Callan:1977pt}, the vacuum
decay rate can be written in the following form:
\begin{equation}
    \gamma=\mathcal A e^{-\mathcal B},
\end{equation}
where $\mathcal B$ is the so-called bounce action and $\mathcal A$ is
a prefactor with mass-dimension four.  Previous tree-level analyses
naively estimated the prefactor $\mathcal A$ based on a typical energy
scale of the bounce.  It has been pointed out that $\mathcal A$ may
deviate significantly from the naive estimation in particular when
there are many particles that couple to the bounce \cite{Endo:2015ixx}
and hence the precise calculation of $\mathcal A$ is important for the
accurate determination of the allowed parameter space. The prefactor
has been first evaluated for the SM in
\cite{Isidori:2001bm} and it has been reevaluated recently with the
correct treatment of zero modes in
\cite{Andreassen:2017rzq,Chigusa:2017dux,Chigusa:2018uuj} using the
prescription proposed in \cite{Endo:2017gal,Endo:2017tsz}. The
prescription has been generalized to a multi-field bounce in
\cite{Chigusa:2020jbn}, which enabled the calculation of precise decay
rates in a more complex setup like the one in this letter. All the coupling constants used below should be
understood as those in the EFT at the renormalization scale of
$Q=M_t$.

The bounce is a spherical object in four-dimensional Euclidean space.
We parameterize the bounce as
\begin{align}
    H=\frac{1}{\sqrt{2}}\mqty(0\\\rho_h(r)),~~~
    \sle{L}=\frac{1}{\sqrt{2}}\mqty(0\\\rho_L(r)),~~~
    \smuon{R}=\frac{1}{\sqrt{2}}\rho_R(r),
\end{align}
where $\rho_I$ ($I=h$, $L$, $R$) are real fields and $r$ is the radius
in the four-dimensional Euclidean space.  Notice that the upper
component of $H$ can be taken to be $0$ without loss of generality due
to the $SU(2)_L\times U(1)_Y$ symmetry.  The directions of the other
fields are chosen such that the trilinear interaction,
$\smuon{R}^\dagger H^\dagger \sle{L}$, becomes non-vanishing.  Then,
the bounce configuration is a solution of the Euclidean equations of
motion:
\begin{equation}
  \partial_r^2\rho_I+\frac{3}{r}\partial_r\rho_I= \pdv{V}{\rho_I},
  \label{EoM}
\end{equation}
satisfying the following boundary conditions:
\begin{align}
    \rho_h(\infty)&=\sqrt{2}v_{\rm EFT},~~~\rho_L(\infty)=\rho_R(\infty)=0,~~~\partial_r \rho_I(0)=0,
\end{align}
where $v_{\rm EFT}$ is the Higgs VEV at the false vacuum in the
EFT.  We obtain the bounce solution by numerically solving
Eq.\ \eqref{EoM} using a modified version of the gradient flow method
\cite{Chigusa:2019wxb, Sato:2019axv,ChiMorSho:Future}.

Next, we explain how we can obtain the prefactor, $\mathcal{A}$, which
takes care of the one-loop effect on the decay rate.  The prefactor is
obtained by the functional determinant of the fluctuation matrix which
is given by the second-order functional derivative of the total action
(containing the total scalar potential given in Eq.\ \eqref{Vtot}).
The prefactor can be expressed as
\begin{equation}
  \mathcal A=2\pi\mathcal J_{\rm EM}\frac{\mathcal{B}}{4\pi^2}
  \mathcal A^{(A,\varphi,c\bar c)}
  \mathcal A^{(\psi)},
\end{equation}
where $A^{(A,\varphi,c\bar c)}$ ($\mathcal{A}^{(\psi)}$) is the effect
of gauge bosons, scalar bosons and Faddev-Popov ghosts (fermions), and
$\mathcal J_{\rm EM}$ is the Jacobian in association with the
zero-mode due to the electromagnetic symmetry breaking.
In calculating $\mathcal{A}$,
we take into account the effects of
the smuons and the Bino
as well as the
$SU(2)_L$ and
$U(1)_Y$ gauge bosons, the Higgs boson, the muons, and the top quark.
$\mathcal A^{(A,\varphi,c\bar c)}$ and $\mathcal A^{(\psi)}$ are given
by the ratios of functional determinants for the partial waves:
\begin{align}
  \mathcal A^{(A,\varphi,c\bar c)}=&\,
  \frac{\det\mathcal M_0^{(c\bar c)}}{\det\mathcal {\widehat M}_0^{(c\bar c)}}\qty(\frac{\det'\mathcal M_0^{(S\varphi)}}{\det\mathcal {\widehat M}_0^{(S\varphi)}})^{-1/2}\qty(\frac{\det'\mathcal M_1^{(SL\varphi)}}{\det\mathcal {\widehat M}_1^{(SL\varphi)}})^{-2}\prod_{\ell=2}^\infty\qty(\frac{\det\mathcal M_\ell^{(SL\varphi)}}{\det\mathcal {\widehat M}_\ell^{(SL\varphi)}})^{-\frac{(\ell+1)^2}{2}},
  \\
  \mathcal A^{(\psi)}=&\,
  \prod_{\ell=0}^\infty\qty(\frac{\det\mathcal M_\ell^{(\psi)}}{\det\mathcal {\widehat M}_\ell^{(\psi)}})^{\frac{(\ell+1)(\ell+2)}{2}},
\end{align}
where the prime indicates the subtraction of zero modes, $\mathcal M_\ell$'s indicate fluctuation matrices around the bounce, and $\widehat{\mathcal M}_\ell$'s indicate
those around the false vacuum.

A general procedure to calculate the decay rate of the false vacuum,
including the prescription for the zero-mode subtraction and the
renormalization, is given in
Refs.\ \cite{Endo:2017gal,Endo:2017tsz,Chigusa:2020jbn}.  We follow
the procedure given in these articles to calculate the decay rate of
the EW vacuum in the model of our interest.  A more detailed
explanation of the calculation of the decay rate of the EW
vacuum in the present model will be given elsewhere \cite{ChiMorSho:Future}.

\section{Numerical results}
\label{sec:results}
\setcounter{equation}{0}

Now we are at the position to show the constraints from the stability
of the EW vacuum.  In order to investigate how large the
slepton mass can be, we do not take into account the constraints from
other considerations, like the collider and dark matter constraints.
These constraints depend on the detail of the model; for example, if
the $R$-parity is violated, these are relaxed considerably.

\begin{figure}[t]
  \centering
  \includegraphics[width=0.65\linewidth]{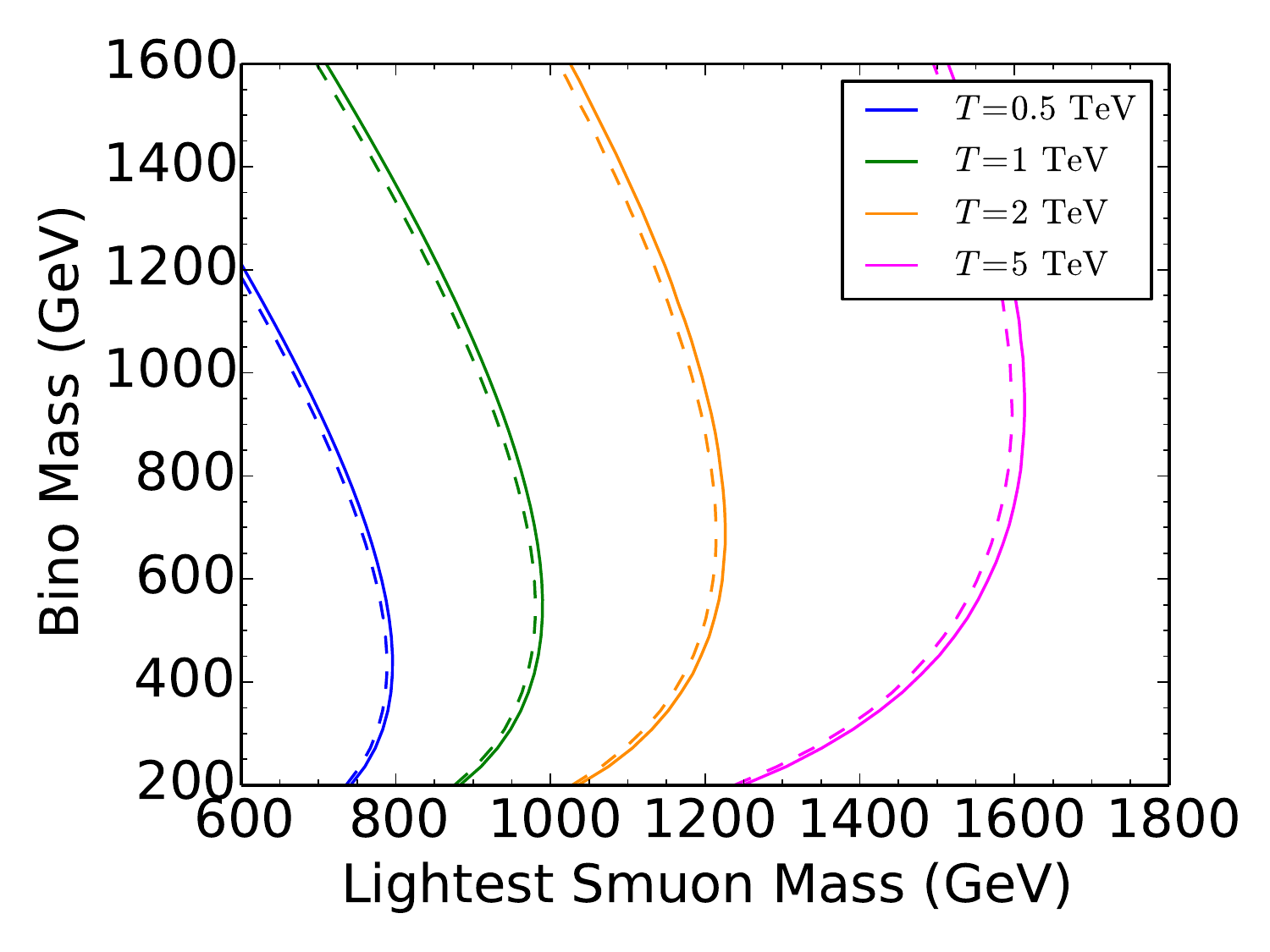}
  \caption{Contours of constant $T$ for the case of $a_\mu^{\rm
      (SUSY)}=25.1\times 10^{-10}$ and $m_R=m_L$. The $\tan\beta$
    parameter is taken to be $10$ (solid) and $50$ (dashed).  The
    blue, green, orange, and magenta lines are for $T=0.5$, $1$, $2$,
    and $5\ {\rm TeV}$, respectively.}
  \label{fig:tparam}
\end{figure}

We first calculate the required value of $T$ to realize a given value
of $a_\mu^{\rm (SUSY)}$ for given values $m_L$, $m_R$, and $M_1$ (as
well as other MSSM parameters).\footnote
{When the Bino mass is relatively large, $|\Delta
  y_{\mu}|$ may become larger than the SM muon Yukawa coupling
  constant $\tilde{y}_\mu$.  In such a case, the EFT muon Yukawa
  coupling constant $y_\mu$ is negative.  (Notice that $\Delta
  y_{\mu}<0$.)  We have checked that our main result,
  Fig.\ \ref{fig:bound}, is unchanged even if we consider only the
  parameter region with $y_\mu>0$.}
In Fig.\ \ref{fig:tparam}, we show the contours of constant $T$
parameter on the $m_{\tilde{\mu}_1}$ vs.\ $M_1$ plane, assuming
$a_\mu^{\rm (SUSY)}=25.1\times 10^{-10}$.  Here we take $m_R/m_L=1$
and $\tan\beta=10$ and $50$.  We can see that the required value of
$T$ to realize $a_\mu^{\rm (SUSY)}\sim \Delta a_\mu$ is insensitive to
the value of $\tan\beta$.  We can also see that the $T$ parameter is
required to be significantly larger than the smuon masses for the case
of heavy sleptons.  Such a choice of $T$, required to solve the muon
$g-2$ anomaly, gives rise to a deeper minimum of the potential in
addition to the EW vacuum.  In such a minimum of the
potential, which we call a charge breaking minimum, the smuons acquire
vacuum expectation values.  The longevity of the EW vacuum is
not guaranteed for the case with the charge breaking minimum.

We calculate the decay rate of the elecroweak vacuum with the
procedure explained in the previous Section.  We parameterize the
decay rare per unit volume as
\begin{align}
  \Seff \equiv - \ln \left( \frac{\gamma}{1\ {\rm GeV}^4} \right).
\end{align}
Then, requiring that the bubble nucleation rate within the Hubble
volume, $\frac{4}{3}\pi H_0^{-3}$, be smaller than $t_{\rm now}^{-1}$,
we obtain
\begin{align}
  \Seff > 386.
  \label{seffbound}
\end{align}

\begin{figure}
  \centering
  \includegraphics[width=0.65\linewidth]{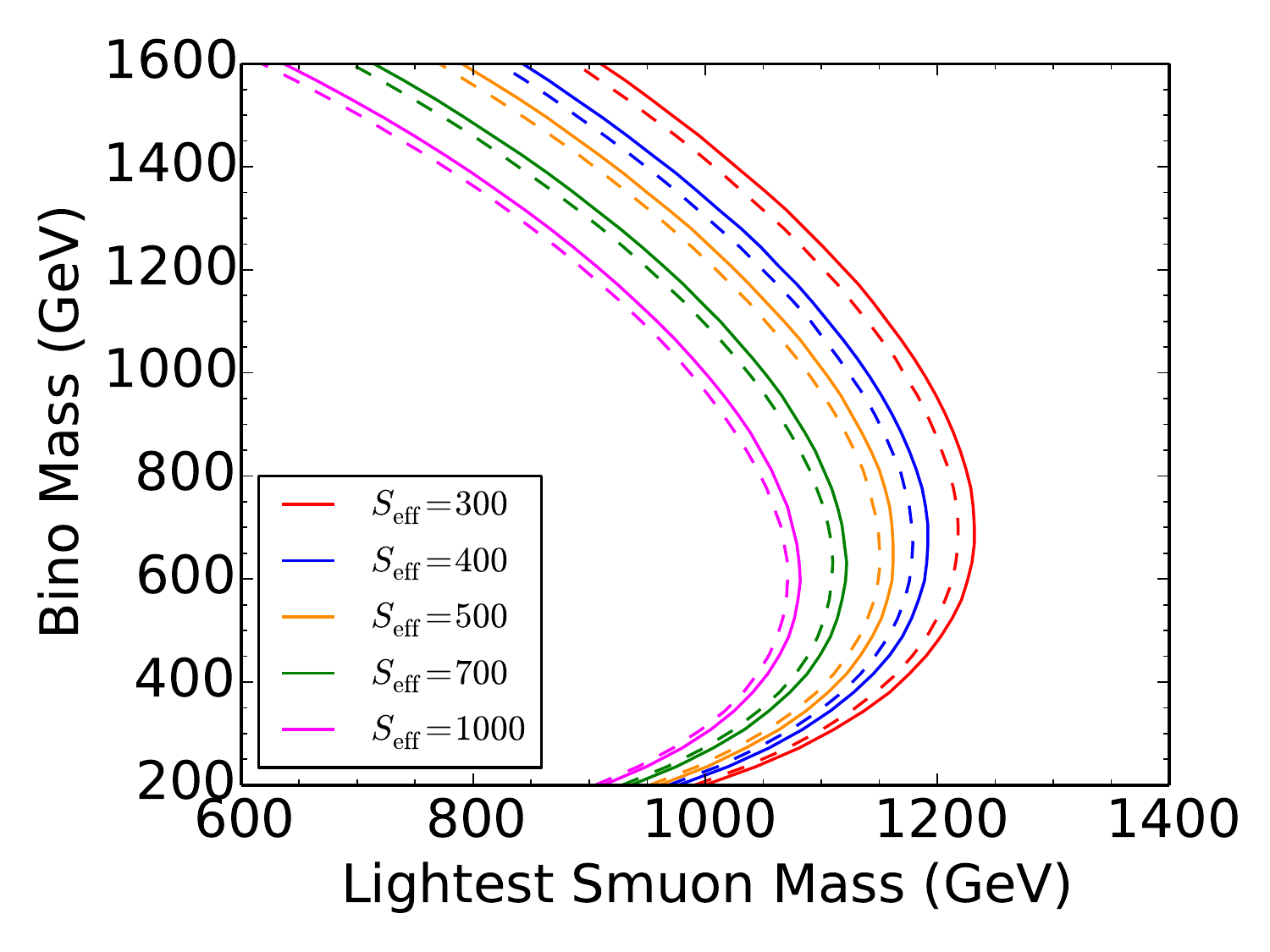}
  \caption{Contours of constant $S_{\rm eff}$, taking $a_\mu^{\rm
      (SUSY)}=25.1\times 10^{-10}$ and $m_R=m_L$.  The red, blue,
    orange, green, and magenta lines are for $S_{\rm eff}=300$, $400$,
    $500$, $700$, and $1000$, respectively.  The solid and dashed
    lines are for $\tan\beta=10$ and $50$, respectively.}
  \label{fig:seff}
  \vspace{5mm}
  \centering
  \includegraphics[width=0.65\linewidth]{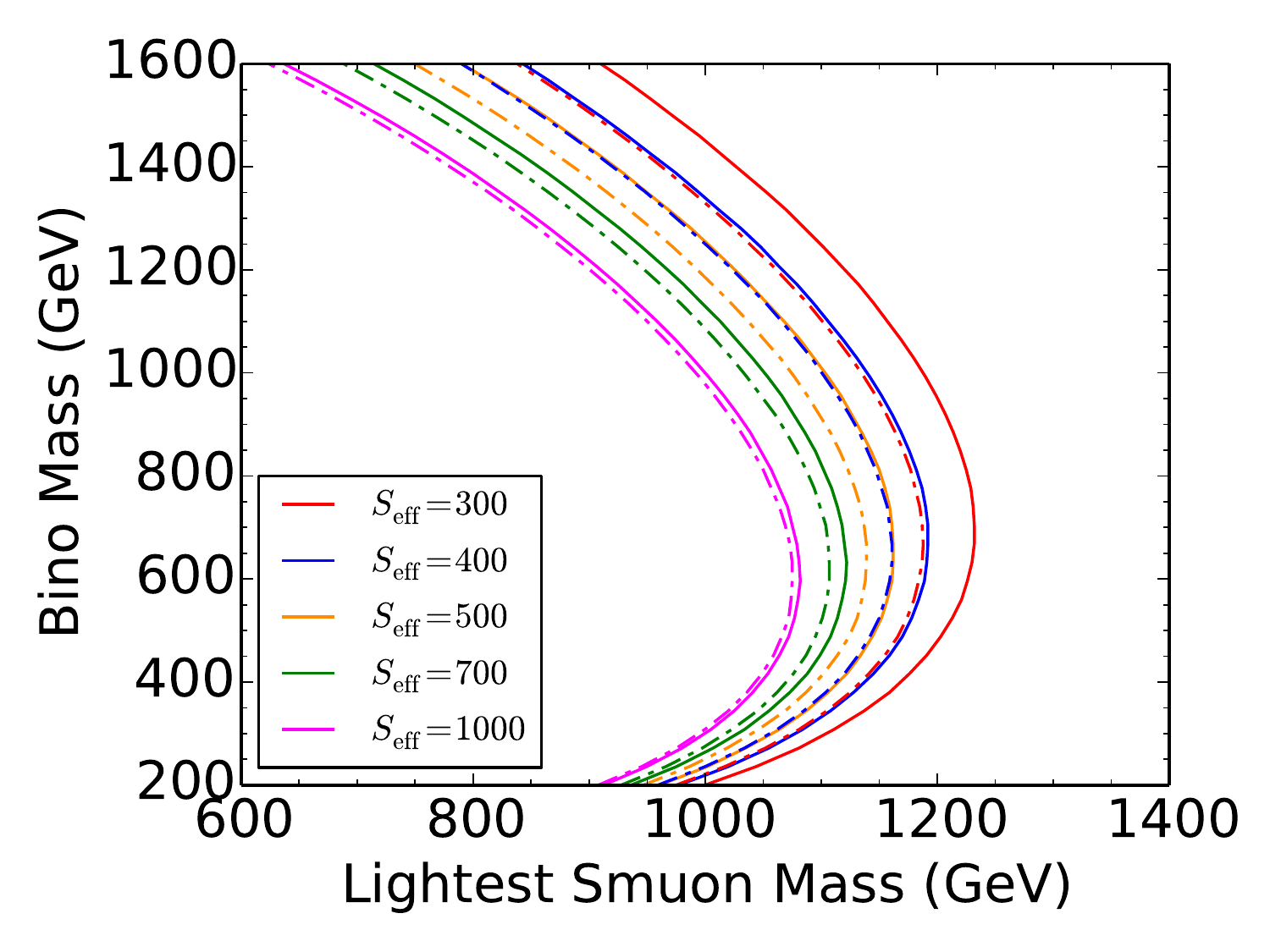}
  \caption{Contours of constant $S_{\rm eff}$ (solid) and $S_{\rm
      eff}^{\rm (tree)}$ (dashdotted), taking $a_\mu^{\rm
      (SUSY)}=25.1\times 10^{-10}$, $m_R=m_L$ and $\tan\beta=10$.  The
    red, blue, orange, green, and magenta lines show the contours on
    which $S_{\rm eff}$ or $S_{\rm eff}^{\rm (tree)}$ is equal to
    $300$, $400$, $500$, $700$, and $1000$, respectively.}
  \label{fig:stree}
\end{figure}

In Fig.\ \ref{fig:seff}, we show the contours of constant $\Seff$ on
the lightest smuon mass vs.\ Bino mass plane with fixing the $T$
parameter by requiring $a_\mu^{\rm (SUSY)}=25.1\times 10^{-10}$; here,
we take $m_R/m_L=1$.  As the lightest smuon becomes heavier, $S_{\rm
  eff}$ becomes smaller and the constraint given in \eqref{seffbound}
may not be satisfied.  Thus, the stability of the EW vacuum gives an
upper bound on the smuon mass assuming that the SUSY contribution is
responsible for the muon $g-2$ anomaly.

In order to see the impact of the one-loop calculation of the
prefactor $\mathcal{A}$, we compare our result with a tree-level
one. For this purpose, because the typical energy scale
of the bounce for the decay of the EW vacuum is often taken
to be around the EW scale, we define
\begin{align}
  \Seff^{\rm (tree)} \equiv \mathcal{B}
  - \ln \left( \frac{v^4}{1\, {\rm GeV^4}} \right).
\end{align}
In Fig.\ \ref{fig:stree}, we show the contours of constant $\Seff$ and
$\Seff^{\rm (tree)}$, taking $a_\mu^{\rm (SUSY)}=25.1\times 10^{-10}$,
$m_L/m_R=1$ and $\tan\beta=10$.  The contours of constant $\Seff$ and
$\Seff^{\rm (tree)}$ show significant deviation.  We find that $\Seff$
and $\Seff^{\rm (tree)}$ differ by $\sim 100$, which results in the
$O(10)\ {\rm GeV}$ difference in the estimation of the upper bound on
the smuon masses.

\begin{figure}
  \centering
  \includegraphics[width=0.65\linewidth]{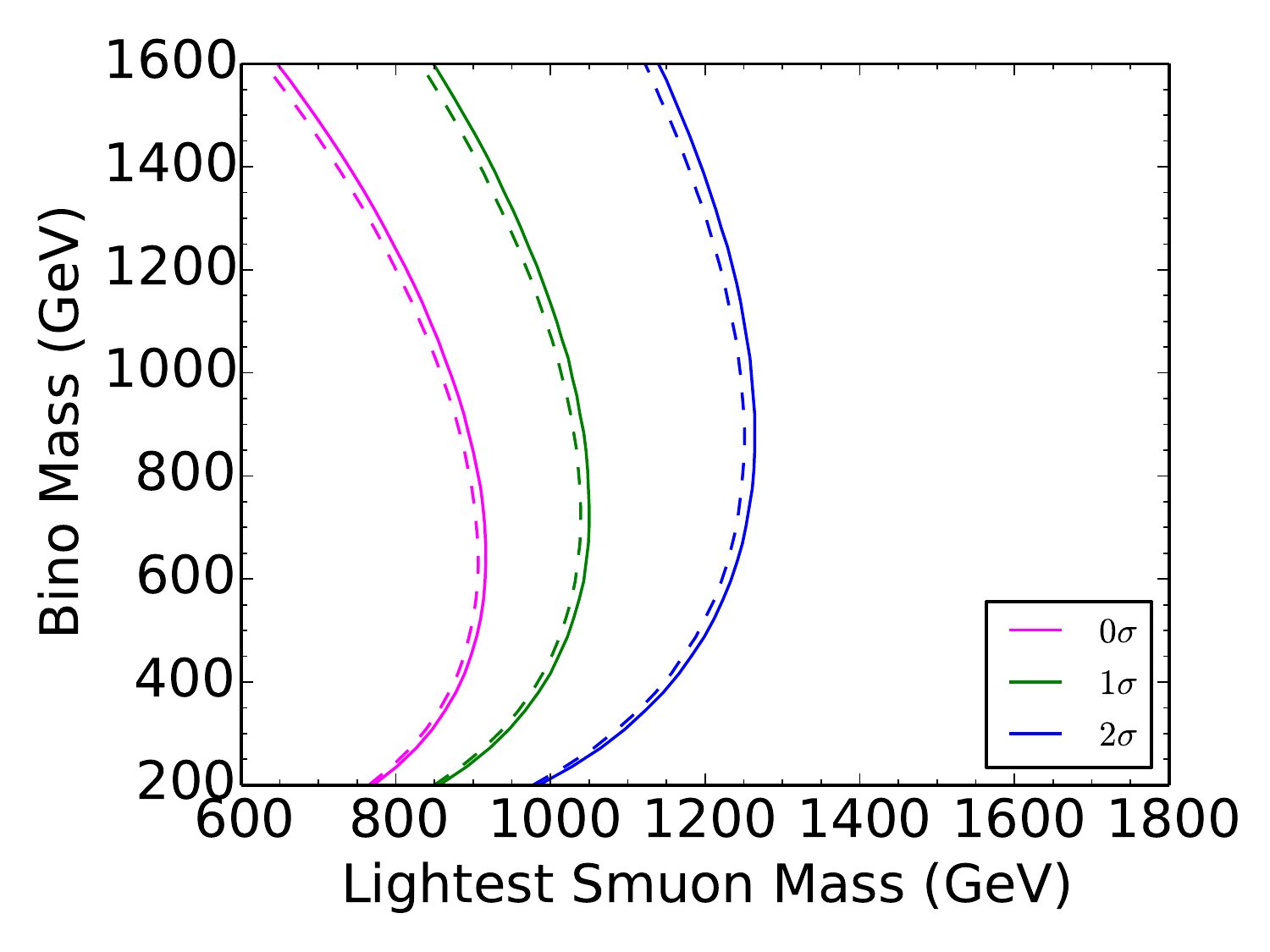}
  \caption{Contours of $\Seff=387$ for $m_R/m_L =0.5$.  The magenta,
    green, and blue lines are for $a_\mu^{\rm (SUSY)}=25.1\times
    10^{-10}$ ($0\sigma$), $19.2\times 10^{-10}$ ($1\sigma$), and
    $13.3\times 10^{-10}$ ($2\sigma$), respectively.  The solid
    (dashed) lines are for $\tan\beta=10$ ($50$).}
  \label{fig:Seff387_r04}
  \vspace{7mm}
  \centering
  \includegraphics[width=0.65\linewidth]{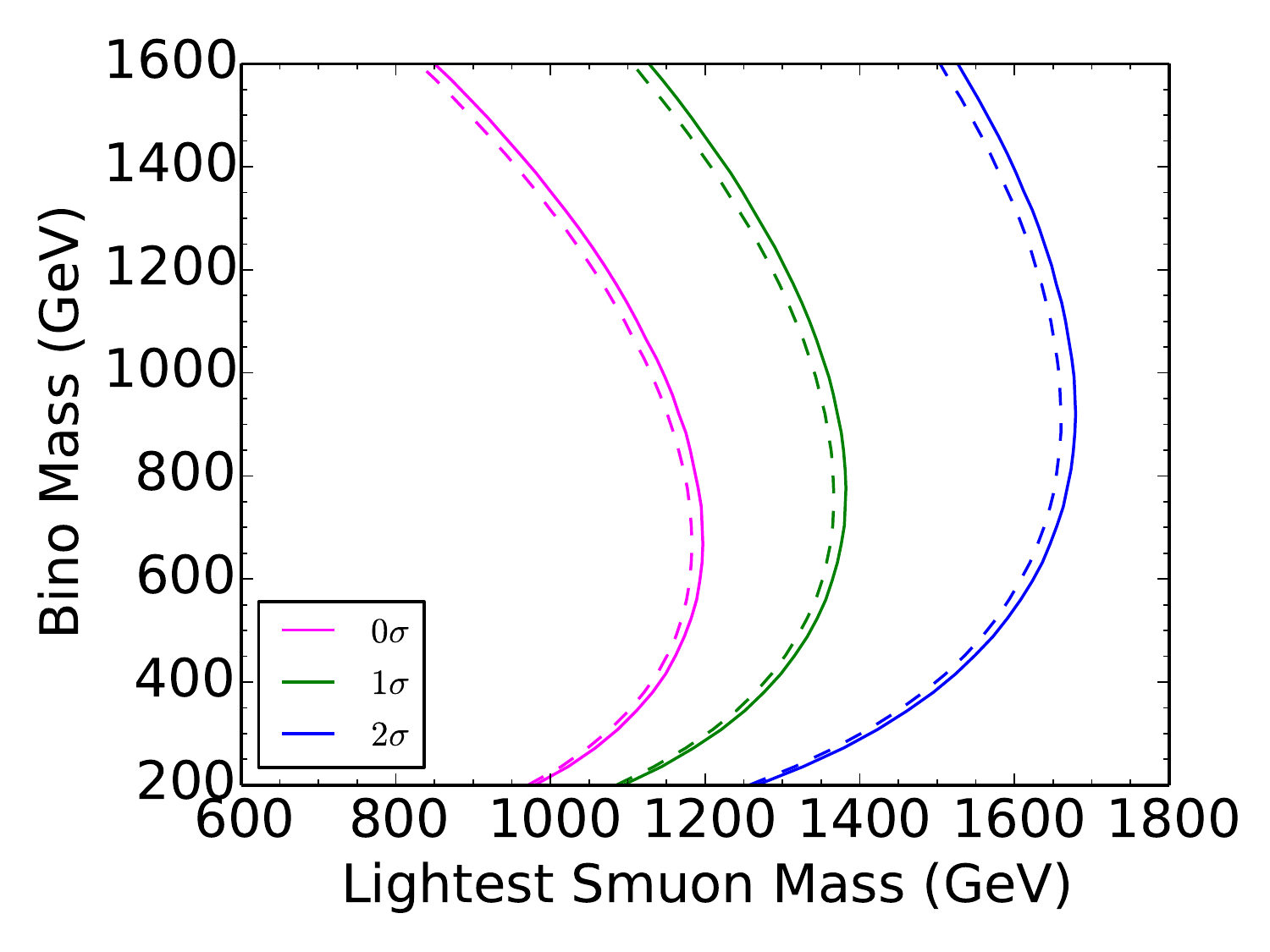}
  \caption{Same as Fig.\ \ref{fig:Seff387_r04}, except
    $m_R/m_L =1$.}
  \label{fig:Seff387_r10}
\end{figure}

\begin{figure}
  \centering
  \includegraphics[width=0.65\linewidth]{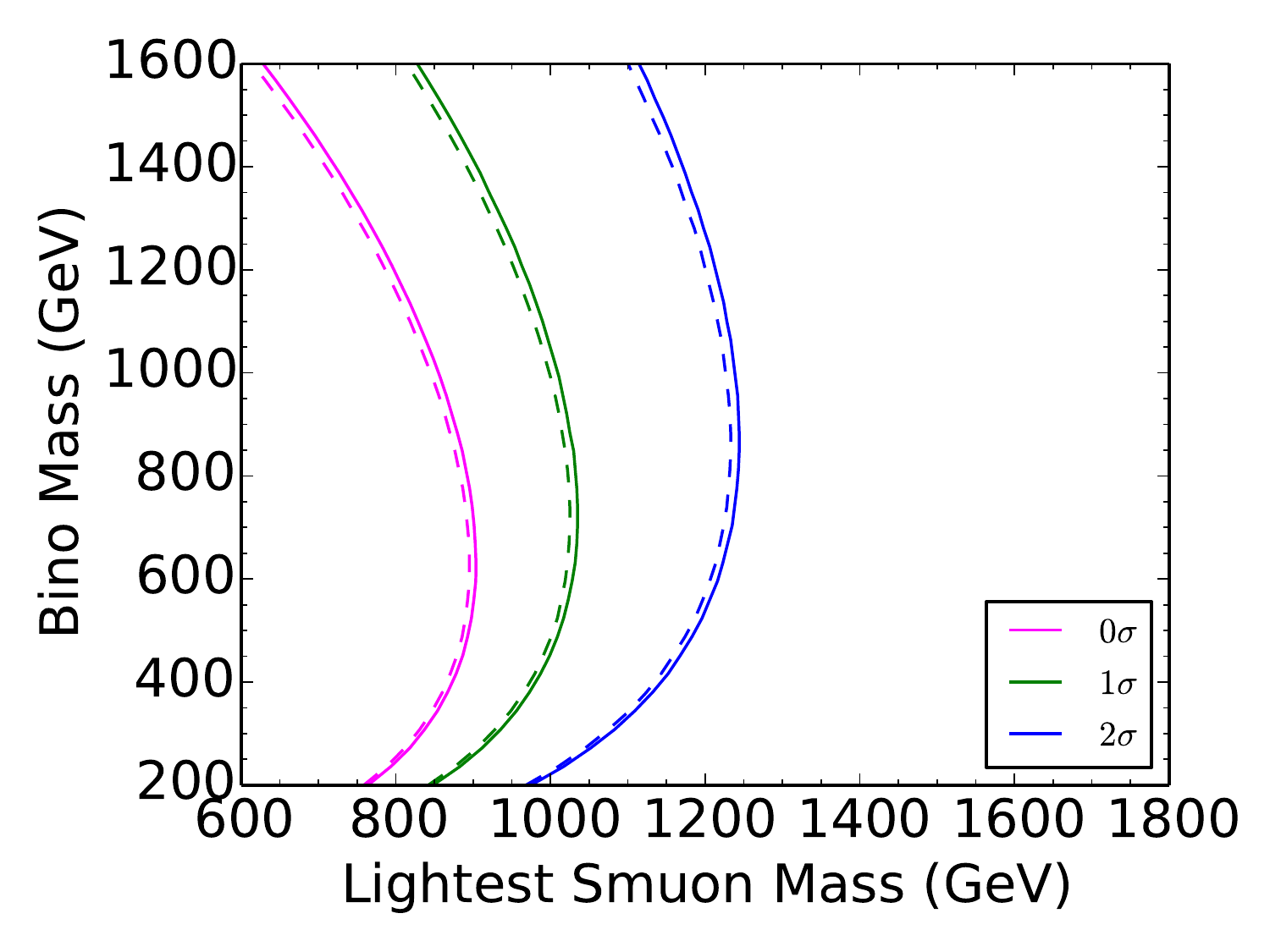}
  \caption{Same as Fig.\ \ref{fig:Seff387_r04}, except
    $m_R/m_L =2$.}
  \label{fig:Seff387_r16}
  \vspace{7mm}
  \centering
  \includegraphics[width=0.65\linewidth]{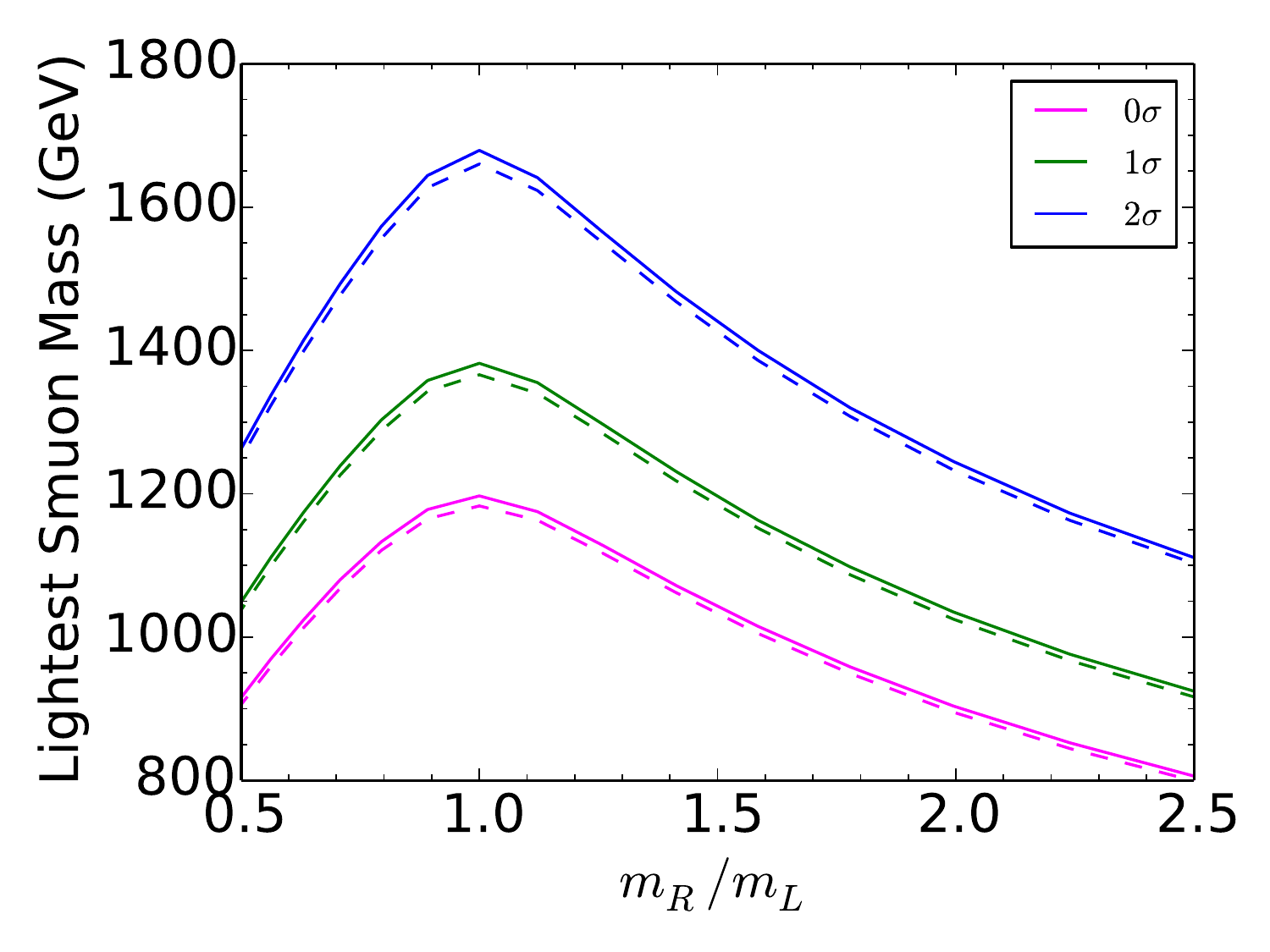}
  \caption{Upper bound on the lightest smuon mass as a function of
    $m_R/m_L$.  The magenta, green, and blue lines are for $a_\mu^{\rm
      (SUSY)}=25.1\times 10^{-10}$ ($0\sigma$), $19.2\times 10^{-10}$
    ($1\sigma$), and $13.3\times 10^{-10}$ ($2\sigma$), respectively.
    The solid (dashed) lines are for $\tan\beta=10$ (50).}
  \label{fig:bound}
\end{figure}

Now, we discuss the constraint on the lightest smuon mass.  In
Figs.\ \ref{fig:Seff387_r04}, \ref{fig:Seff387_r10}, and
\ref{fig:Seff387_r16}, we show the contours of $\Seff=386$ for
$m_R/m_L =0.5$, $1$, and $2$, for $a_\mu^{\rm (SUSY)}=25.1\times
10^{-10}$ ($0\sigma$), $19.2\times 10^{-10}$ ($1\sigma$), and
$13.3\times 10^{-10}$ ($2\sigma$).  Requiring that $a_\mu^{\rm
  (SUSY)}$ is comparable to $\Delta a_\mu$, we can see that the
lightest smuon mass becomes maximally large when the Bino mass is
$\sim 0.5-1\ {\rm TeV}$.  In addition, as expected, the upper bound on
the smuon mass becomes larger as $a_\mu^{\rm (SUSY)}$ becomes smaller.
Notice that our smuon mass bound for the case of $m_R/m_L =1$ is close
to the one given in Ref.\ \cite{Endo:2021zal}, which is based on the
tree-level estimation of the decay rate.

Varying the Bino mass, we determined the maximal possible value of the
lightest neutralino mass for fixed values of $\tan\beta$ and
$a_\mu^{\rm (SUSY)}$.  The result is shown in Fig.\ \ref{fig:bound},
in which the upper bound on the lightest neutralino mass is given as a
function of the ratio $m_R/m_L$. We can see that the upper bound
becomes the largest when $m_R\simeq m_L$.  Requiring $a_\mu^{\rm
  (SUSY)}=25.1\times 10^{-10}$ ($0\sigma$), $19.2\times 10^{-10}$
($1\sigma$), and $13.3\times 10^{-10}$ ($2\sigma$) with $\tan\beta=10$
($50$) and $m_R= m_L$, the lightest smuon mass is required to be
smaller than $1.20$, $1.38$ and $1.68\ {\rm TeV}$ ($1.18$, $1.37$
and $1.66\ {\rm TeV}$), respectively.  The bound is insensitive to the
choice of $\tan\beta$.  The muon $g-2$ anomaly can be hardly explained
by the MSSM contribution if the lightest smuon is heavier than this
bound.

\section{Conclusions and discussion}
\label{sec:conclusions}
\setcounter{equation}{0}

In this letter, we have studied the stability of the EW
vacuum in the MSSM, paying particular attention to the parameter
region where the muon $g-2$ anomaly can be explained by the SUSY
contribution.  We consider the case where the Higgsino mass parameter
$\mu$ is significantly large; in such a case, the SUSY contribution to
the muon $g-2$ is enhanced so that the muon $g-2$ anomaly can be
explained with relatively large values of the smuon masses.  With $\mu$
being large, however, the smuon-smuon-Higgs trilinear coupling is
enhanced, and there may show up a charge breaking minimum of
the potential, resulting in the meta-stability of the EW
vacuum.  With the size of the SUSY contribution to the muon $g-2$
being fixed to alleviate the muon $g-2$ anomaly, the trilinear
coupling is more enhanced with a larger value of the smuon mass.  Thus,
if the smuon mass is too large, the muon $g-2$ anomaly cannot be
solved in the MSSM even if we consider a very large value of $\mu$
because the longevity of the EW vacuum cannot be realized.

We have performed a detailed calculation of the decay rate of the
EW vacuum, assuming that the SUSY contribution to the muon
anomalous magnetic moment is large enough to alleviate the muon $g-2$
anomaly.  Our calculation is based on the state-of-the-art method to
calculate the decay rate of the false vacuum, which includes the
one-loop effects due to the field coupled to the bounce.  The most
important advantage of the inclusion of the one-loop effect is to
determine the mass scale of the prefactor $\mathcal{A}$, which is
mass-dimension $4$.  Another advantage is that the scale dependence of
the bounce action $\mathcal{B}$ can be canceled by that of
$\mathcal{A}$ at the leading-log level.  Requiring $a_\mu^{\rm
  (SUSY)}=25.1\times 10^{-10}$ ($0\sigma$), $19.2\times 10^{-10}$
($1\sigma$), and $13.3\times 10^{-10}$ ($2\sigma$), we found that the
lightest smuon should be lighter than 
$1.20$, $1.38$ and $1.68\ {\rm TeV}$ 
($1.18$, $1.37$ and $1.66\ {\rm TeV}$)
for $\tan\beta=10$
($50$), respectively.  It is challenging to find such a heavy smuon
with collider experiments.  A very high energy collider, like muon
colliders \cite{Delahaye:2019omf}, the FCC \cite{Mangano:2016jyj,
  Contino:2016spe, Golling:2016gvc}, or the CLIC \cite{CLICdp:2018cto}
may be able to perform a conclusive test of the SUSY interpretation of
the muon $g-2$ anomaly.

In this letter, we assumed that the superparticles other than the
smuons and the Bino are so heavy that they are irrelevant for the muon
$g-2$ as well as for the stability of the EW vacuum.  If some
of the superparticles are as light as the smuons and the Bino, the
upper bound on the smuon mass we obtained may become more stringent.
For example, if the stau is relatively light, then the decay rate of
the EW vacuum may become larger because the large $\mu$ also
enhances the stau-stau-Higgs trilinear coupling which is orders of
magnitude larger than the smuon-smuon-Higgs coupling.  In such a case,
the upper bound on the slepton mass becomes more stringent compared to
the case only with the smuons.  More detailed discussion on such a
case will be given elsewhere \cite{ChiMorSho:Future}.

\vspace{2mm}
\noindent{\it Acknowledgments:} S.C. is supported by JSPS KAKENHI
Grant No.\ 20J00046, and also by the Director, Office of Science,
Office of High Energy Physics of the U.S. Department of Energy under
the Contract No.\ DE-AC02-05CH1123.
T.M. is supported by JSPS KAKENHI Grant Nos.\ 16H06490 and 18K03608.
Y.S. is supported by I-CORE Program of the
Israel Planning Budgeting Committee (grant No.\ 1937/12). 
The authors gratefully acknowledge the computational and data resources provided by the Fritz Haber Center for Molecular Dynamics.

\bibliographystyle{jhep}
\bibliography{ref}

%%%%%%%%%%%%%%%%%%%%%%%%%%%%%%%%%%%%%%%

\end{document}